\documentclass[preprint,superscriptaddress,showpacs,prb,amsmath,amssymb,endfloats*]{revtex4}

\usepackage{graphicx}
\usepackage{dcolumn}
\usepackage{bm}

\usepackage{psfig}

\begin{document}

\title{Neutron Scattering and magnetization studies of Ba$_2$Cu$_{2.95}$Co$_{0.05}$O$_4$Cl$_2$:
A decorated two-dimensional antiferromagnet
}

\author{M.K.~Ramazanoglu} 
\affiliation{Department of Physics, University of Toronto, Toronto, 
Ontario, M5S 1A7, Canada}

\author{P.S.~Clegg}
\affiliation{Department of Physics, University of Toronto, Toronto, 
Ontario, M5S 1A7, Canada}
\affiliation{SUPA, School of Physics, University of Edinburgh, Edinburgh EH9 3JZ, UK}

\author{S.~Wakimoto} 
\affiliation{Department of Physics, University of Toronto, Toronto, 
Ontario, M5S 1A7, Canada}
\affiliation{JAERI, Advanced Science Research Center, Tokai, Ibaraki 319-1195, Japan}

\author{R.J.~Birgeneau} 
\affiliation{Department of Physics, University of Toronto, Toronto, 
Ontario, M5S 1A7, Canada}
\affiliation{Department of Physics, University of California, Berkeley, California 94720}

\author{S.~Noro}
\affiliation{Graduate School of Integrated Science, Yokohama City University, Seto
22-2, Kanazawa-ku, Yokohama 236, Japan}

\date{\today}

\begin{abstract}
Ba$_2$Cu$_3$O$_4$Cl$_2$ has two inter-penetrating square Cu sublattices, one with 
square root 2 times the in-plane spacing of the other. Isotropic magnetic interactions between the
two sublattices are completely frustrated. Quantum fluctuations
resolve the intrinsic degeneracy in the ordering direction of the more
weakly coupled sublattice in favor of collinear ordering. 
We present neutron scattering and magnetization studies of the magnetic
structure when the Cu ions are substituted with Co. The Co spins create new magnetic
interactions between the two sublattices. The ordering behavior of both Cu sublattices is
retained largely unmodified. Between the phase transitions of the two sublattices spin-glass
behavior is observed. Magnetization results show a strong enhancement to the ferromagnetic aspect of the
magnetic structure. The combination of glassy behavior and large moments strongly suggest that the Co moments
induce the formation of local canted states.
\end{abstract}


\maketitle

\section{Introduction}
\label{sec:introduction}

A wealth of experiments have explored the competition between different types of
fluctuation which disrupt ordered phases~\cite{Frisken1992, Chan1996, Birgeneau1998, Clegg2005}. 
A typical combination is thermal fluctuations
and quenched random fluctuations due to disorder. The strength of the thermal
fluctuations is controlled via the temperature while the strength of the quenched
fluctuations is controlled by dilution of the constituents. Since both forms of fluctuation perturb the order
they are never in profound competition. Here, by contrast, we study the interaction between quantum fluctuations,
which give rise to an order-from-disorder transition~\cite{Shender1982}, 
and quenched random fluctuations. As the quantum
fluctuations drive the ordering in the chosen material the two sets of fluctuations are in direct competition and
the results can be expected to be fascinating.

Ba$_2$Cu$_3$O$_4$Cl$_2$ is a lamellar cuprate with two Cu sublattices~\cite{Yamada1995}. 
The Cu$_I$
sublattice has CuO$_2$ layers and orders antiferromagnetically at $T_{N,I} = 332.4 \pm 0.8$~K. The
Cu$_{II}$ sublattice interpenetrates the Cu$_I$ sublattice with one Cu$_{II}^{2+}$
ion at the center of every second plaquette. The position of the Cu$_{II}$ sublattice relative
to a Cu$_I$ plaquette is shown in Fig.~\ref{coupling}(a). 
The plaquettes are staggered from one plane
to the next. The Cu$_{II}$ moments order antiferromagnetically at $T_{N,II} = 31.31 \pm 0.01$~K. The
isotropic couplings between the Cu$_I$ and Cu$_{II}$ sublattices are completely
frustrated and this means that more subtle interactions are manifest~\cite{Chou1997}. 
Collinear
ordering of the second sublattice with respect to the first is preferred due to the
contribution of quantum fluctuations. As a result the Cu$_{II}$ system has Ising
class critical properties.

Detailed studies of the pure \textit{2342} materials have been carried out in recent years~\cite{Yamada1995,
Chou1997, Noro2002, Kim1999, Kim2001, Kim2001b}.
In an applied magnetic field, in the temperature range $T_{N,II} < T < T_{N,I}$, Sr$_2$Cu$_3$O$_4$Cl$_2$ develops a
small magnetic moment on the Cu$_{II}$ sites. Chou and coworkers studied the behavior as a function of
temperature, field strength and field orientation~\cite{Chou1997}.
Combined with the known behavior of the Cu$_I$
order parameter it was possible to determine the nature of the coupling between Cu$_I$ and
Cu$_{II}$ moments. The isotropic average coupling $J_{av} = -12 \pm 9$~meV and may originate
from superexchange and direct exchange. The anisotropic coupling is $|J_{pd}| = 27 \pm 1~\mu$eV
and is pseudo-dipolar.
Noro and coworkers studied the weak ferromagnetism 
in Ba$_2$Cu$_3$O$_4$Cl$_2$ in the pure material and with Zn or Ni substitution~\cite{Noro2002}. The behavior of the pure material
has similar magnetization, ferromagnetism and susceptibility to the Sr material. Neither substitution of
the spin system resulted in any new phase transitions or glassy behavior. Adding Zn reduced
the ordered moment by a small amount. Adding Ni reduced the ordered moment by almost a factor of
2 without altering the transition temperatures. Apart from very close to the two ordering
transitions the ferromagnetic component is entirely suppressed. Furthermore the anisotropy in the
in-plane susceptibility disappears. These effects suggest that Ni substitution weakens the
coupling between the two sublattices.

Kim and coworkers~\cite{Kim1999,Kim2001} 
made a comprehensive elastic, quasi-elastic and inelastic neutron scattering
study of Sr$_2$Cu$_3$O$_4$Cl$_2$. The role of quantum fluctuations in coupling Cu$_I$ and Cu$_{II}$ subsystems
were demonstrated via the dramatic variation in the energies of the spin-wave modes as the
Cu$_{II}$ subsystem orders. The behavior is in quantitative agreement with theoretical
predictions~\cite{Harris2001}. The dispersion of a Cu$_{II}$ mode was measured out to the zone boundary and this
provided precise verification of theoretical calculations.
The major interactions were found and are indicated in
Fig.~\ref{coupling}(a). For the Sr material $J_I \sim 130$~meV while 
$J_{II} \sim |J_{I-II}| \sim 10$~meV ($J_I$ and $J_{II}$ had not been determined by Chou and coworkers~\cite{Chou1997}).
The similar transitions and transition temperatures suggest 
that the coupling strengths will be very similar for the Ba material.

Here we are studying this intriguing two sublattice antiferromagnet with the Cu moments mixed
with $\sim 2\%$ Co. This is seen to modify some of the magnetic behavior substantially. Similar
magnetic systems have been considered theoretically and were called a \textit{decorated} model.
Villain considered whether spin-glass behavior could occur for an insulator with only
antiferromagnetic interactions~\cite{Villain1979}. He first demonstrated that competing ferromagnetic and
antiferromagnetic interactions emerge for a two sublattice Ising antiferromagnet with spin
dilution of one of the sublattices. Subsequently a number of promising scenarios were considered
including local canted states: An antiferromagnetic ground state is perturbed when an interstitial
magnetic impurity is introduced. The neighboring moments cant due to their interaction with the
interstitial moment and this produces a local canted state. The local canted state is characterized by an (n-1)
component magnetic moment which is perpendicular to the ordering direction of the
antiferromagnetic ground state. If there are many local canted states in a system they will interact and
undergo a spin-glass transition. Hence different components of the magnetic moment order
separately via antiferromagnetic and spin-glass transitions.

Using an alternative approach, Henley considered the role of thermal, 
quantum and concentration fluctuations in a frustrated
antiferromagnet~\cite{Henley1987, Henley1989}. 
He considered a two sublattice antiferromagnet, in contrast to Ba$_2$Cu$_3$O$_4$Cl$_2$, both
sublattices were on the same sized lattice. He, following others~\cite{Shender1982}, 
argued that thermal and quantum
fluctuations of one sublattice give rise to collinear ordering in the other. This is because the
largest fluctuations are transverse to the spin direction for both sublattices. The coupling
between these fluctuations reduces the energy. The fluctuations in composition change the
magnitude of the magnetic moment at random and this is equivalent to a longitudinal fluctuation.
In order to couple transverse fluctuations to these longitudinal fluctuations the neighboring
spins on the other sublattice align perpendicular to them. If the perturbation due to variations
in composition is too small the thermal and quantum fluctuations will maintain the Ising type symmetry. The
composition fluctuations will then give random exchange fields which will pin the ordering
direction in various locations and may prevent the formation of a long-range ordered state.

Experimental studies of Ba$_2$Cu$_{2.95}$Co$_{0.05}$O$_4$Cl$_2$ are described and the results presented
(Sec.~\ref{sec:results}) and
discussed (Sec.~\ref{sec:discussion}) below.

\section{Results}
\label{sec:results}

\begin{figure}
\includegraphics[scale=0.6]{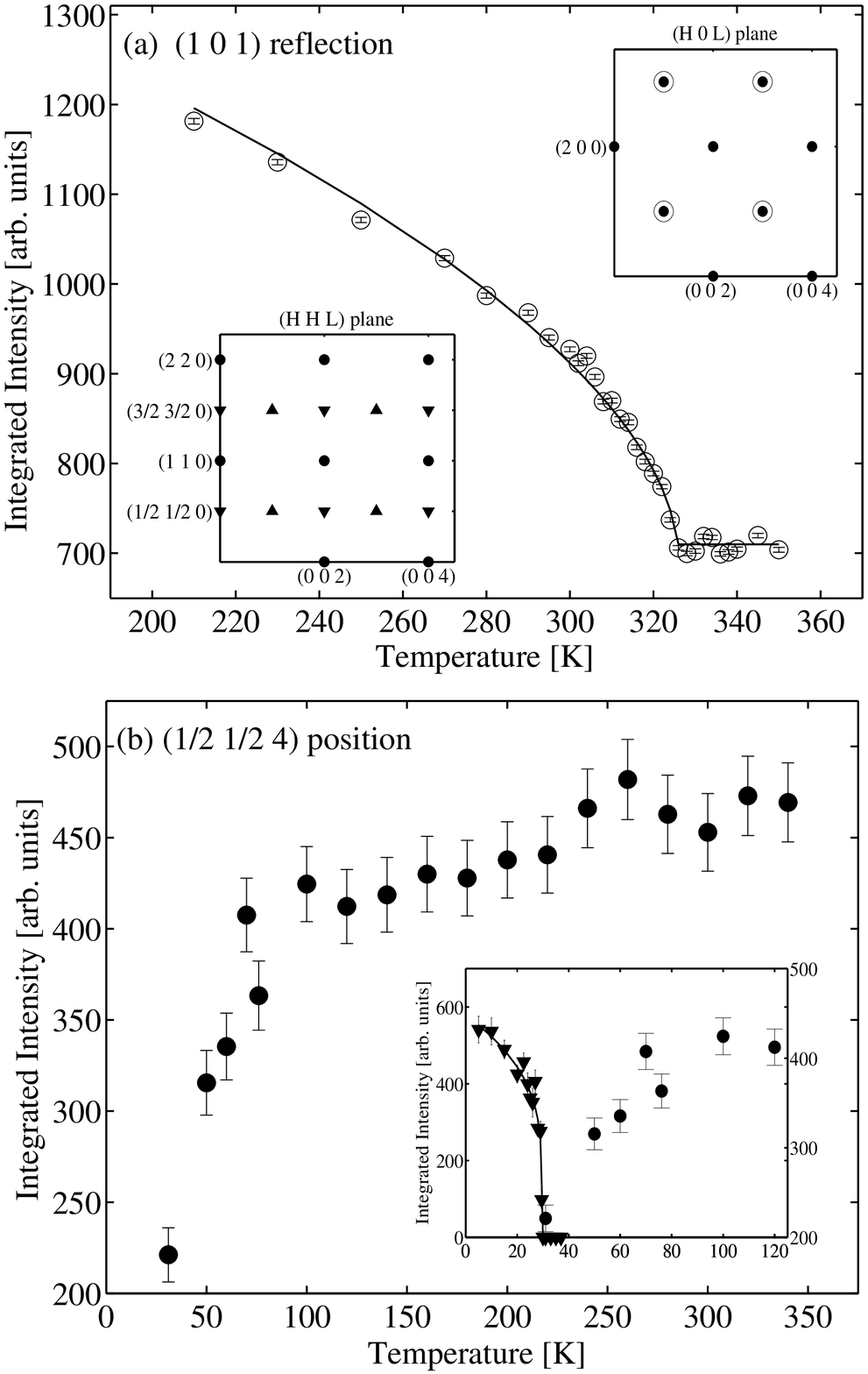} 
\caption{ \label{Tdependence} 
(a) The temperature dependence of the scattered intensity at the (1~0~1) position.
The growth in intensity corresponds to the ordering of the moments on the Cu$_I$ sublattice. The
solid line is the known behavior of the pure material~\cite{Yamada1995}. Inset: reciprocal lattice diagrams of the (H~0~L)
and (H~H~L) planes explored in these experiments. The full circles are the nuclear Bragg reflections; the open circles are
the Cu$_I$ magnetic Bragg reflections and the full (up and down) triangles are the magnetic Bragg reflections from the two
Cu$_{II}$ domains.
(b) The temperature dependence of the short-range correlations at the
($\frac{1}{2}$~$\frac{1}{2}$~$4$) position. The intensity drops for $T \leq 100$~K. 
Inset: the triangles and the circles show the
temperature dependence of the ($\frac{1}{2}$~$\frac{1}{2}$~$2$) and ($\frac{1}{2}$~$\frac{1}{2}$~$4$)
positions respectively. The growth of intensity at the ($\frac{1}{2}$~$\frac{1}{2}$~$2$) position is
due to antiferromagnetic order involving the Cu$_{II}$ sites.}
\end{figure}

A single crystal of Ba$_2$Cu$_{3-x}$Co$_x$O$_4$Cl$_2$ with $x = 0.05$
was prepared by a direct melt
method from powder samples~\cite{Noro2002}. The laminar single crystal had dimensions
$5 \times 4 \times 0.8$~mm$^3$ and a well
developed (0 0 1) plane. The crystal is tetragonal~\cite{Kipka1976} (space group \textit{I4/mmm})
with lattice constants a~=~5.514~\AA~and 
c~=~13.711~\AA~at
T~=~30~K. The Cu$_I$ ordering was studied using neutron scattering with the crystal mounted in the (H~0~L) plane while the
Cu$_{II}$ magnetism was observed in the (H~H~L) scattering plane. The location of reflections in these planes is shown
inset to Fig.~\ref{Tdependence}(a).
The neutron scattering measurements were performed on the BT7 and BT9 instruments at
the National Institute for Standards and Technology, Center for Neutron Research (NCNR). 
The incident neutrons of energy 14.7~meV were selected with a (0~0~2)
graphite monochromator and the collimation was generally [60'-40'-40'-80'] on BT7 and
[40'-40'-40'-80'] on BT9. A pyrolytic graphite filter
was used to suppress higher order contamination. An (0~0~2) graphite analyzer was
used to improve the instrumental resolution and to suppress background counts. For
the quasi-elastic study of the diffuse scattering a two-axis configuration was
employed.
The magnetization was measured using a SQUID magnetometer over a temperature range 4 to
400~K.

The temperature dependence of the neutron scattering intensity of the (1~0~1) reflection is shown in 
Fig.~\ref{Tdependence}(a). 
The additional component to the 
scattering below $T = 324.9 \pm 0.4$~K is due to the Cu$_I$ antiferromagnetic structure. The
solid line shows the order parameter behavior for pure Ba$_2$Cu$_3$O$_4$Cl$_2$. For this line the exponent
$\beta = 0.29 \pm 0.02$ and this is close to, but somewhat less than, the value $\beta = 0.35$ expected
for a 3D XY phase transition.
The addition of Co is seen to have very little effect on this transition.

\begin{figure}
\includegraphics[scale=0.65]{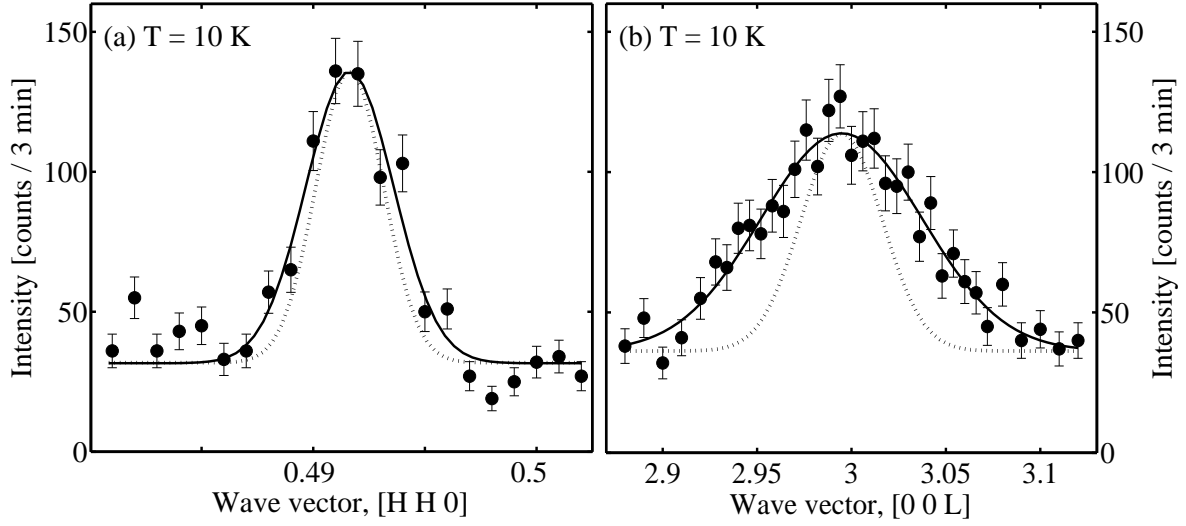} 
\caption{ \label{two-d} Scans of neutron scattering intensity versus wave-vector transfer for
(a) $\mathbf{q}$ = (H~H~3) and (b) $\mathbf{q}$ = ($\frac{1}{2}$~$\frac{1}{2}$~L). The reflection
corresponds to the antiferromagnetic order of the Cu$_{II}$ sublattice. The dotted line is the
instrumental resolution. Correlation between the Cu$_{II}$ ordering on different planes is
evidently short range.}
\end{figure}

New behavior occurs at lower temperatures involving the Cu$_{II}$ spins. Figure~\ref{Tdependence}(b) 
shows the neutron scattering observed at the 
($\frac{1}{2}$~$\frac{1}{2}$~$4$) position for temperatures above $30$~K. The choice of a large $L$ component to the
wave vector enhances the scattering from magnetic moments lying within the planes. The scattering was weak
with no clear peak
and the results presented are integrated counts. The measurements were made in
two-axis configuration in order to reduce the energy resolution. The
intensity is observed to decrease slowly with decreasing temperature with a more
substantial drop beginning around $T \sim 100$~K. This diffuse scattering is different in character to that observed in the
pure material~\cite{Kim2001, Kim2001b}. 
In the pure case rods of scattering are observed in the vicinity of $T_{N,II}$ that reach a peak intensity
at the transition and diminish in the ordered phase. The diffuse scattering observed here corresponds to very
short-range correlations which appear to freeze out well above $T_{N,II}$.
The results
were confirmed to be similar via more sparse measurements 
at the ($\frac{1}{2}$~$\frac{1}{2}$~$2$) position. The addition of 2\% of Co increases the effective interaction
strength between the Cu$_{II}$ spins. One quarter of the Cu$_{II}$ moments have a Co as
a nearest neighbor on one of the two sublattices. For these moments the \textit{transition temperature}
effectively triples - in line with the diffuse scattering.
Inset to
Fig.~\ref{Tdependence}(b) 
is the scattering measured at the ($\frac{1}{2}$~$\frac{1}{2}$~$2$)
position below $T = 30$~K superimposed on the temperature dependence of the
diffuse scattering. The enhanced intensity of scattering results from the
ordering of the Cu$_{II}$ moments. In this case $\beta = 0.102 \pm 0.011$. 
As seen here the temperature dependence of the
order parameter is very similar to that of the pure material~\cite{Yamada1995} where $\beta =
0.117 \pm 0.007$. Both of these are close to the 2D Ising value of $\beta = 0.125$.
Figure~\ref{two-d} demonstrates the short-range nature of the low temperature
magnetic structure perpendicular to the planes. Within the
Cu$_3$O$_4$ planes the ($\frac{1}{2}$~$\frac{1}{2}$~$3$) peak width is comparable 
to the instrumental
resolution. By contrast the
magnetic reflection is seen to be broader than the instrumental resolution out of the
plane. The magnetism is correlated over $157 \pm 14$~\AA~ in the [0~0~L] 
direction which is a little over ten unit cells.
This result suggests that two-dimensional Ising behavior is retained.

\begin{figure}
\includegraphics[scale=0.55]{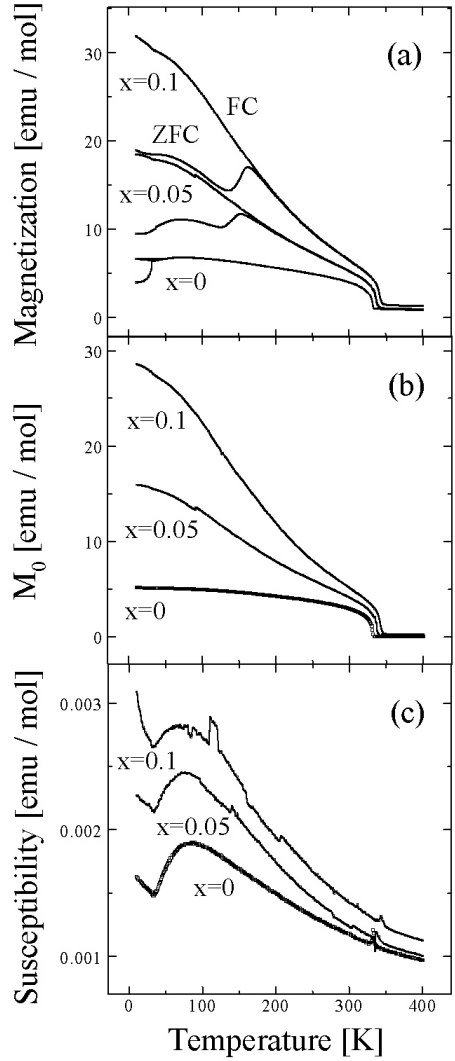} 
\caption{\label{Noro} Temperature dependence of (a) the magnetization along [1~1~0] in
Ba$_2$Cu$_{3-x}$Co$_x$O$_4$Cl$_2$ single crystals, measured after field cooling (0.1\,T)
and zero-field cooling. The magnetic hysteresis shows the presence of the
short-range order.
(b) the weak ferromagnetic moment M$_0$ along [1~1~0]  in
Ba$_2$Cu$_{3-x}$Co$_x$O$_4$Cl$_2$ single crystals. The canting angle of Cu$_{II}$ and Cu$_I$ 
moments increases
steeply due to the doping of Co ions.
(c) the magnetic susceptibility along [1~1~0] in
Ba$_2$Cu$_{3-x}$Co$_x$O$_4$Cl$_2$  single crystals. The increase in the paramagnetic
susceptibility by Co doping is associated with the formation of local canted states.}
\end{figure}

Magnetization measurements have been carried out for Ba$_2$Cu$_3$O$_4$Cl$_2$ and for two levels
of Co substitution (x~=~0.05,~0.1). One of these is similar to the sample described above.
Figure~\ref{Noro}(a) shows the magnetization along the [1~1~0] direction.
The two traces for each sample are for zero-field cooling  (ZFC)
and cooling in a 0.1\,T field (FC). All samples exhibit a transition at $T \sim 330$\,K
corresponding to ordering on the Cu$_I$ sites. The transition shifts up in temperature
with Co substitution due to the increase in the average moment. For the pure Ba$_2$Cu$_3$O$_4$Cl$_2$
sample the ZFC curve shows a sharp discontinuity at $T \sim 30$\,K corresponding to the onset
of antiferromagnetic order. The key feature of these data is the evidence of a transition at
intermediate temperature induced by the Co substitution.
For x~=~0.05 the first transition occurs at $T \sim 150$\,K and
appears rounded; a second discontinuity is observed
at $T \sim 30$\,K. These results suggest that order is occurring on the Cu$_{II}$ sites
in two steps. The magnetization observed is greatly enhanced as a result of substitution of a
small amount of Co.

The dependence of magnetization on the applied field strength can be fitted by 
$M(H) = M_0 + \chi H$, indicating the existence of a weak ferromagnetic component $M_0$. The
temperature dependence of $M_0$ is shown in Fig.~\ref{Noro}(b) and the susceptibility in
Fig.~\ref{Noro}(c). The results for pure Ba$_2$Cu$_3$O$_4$Cl$_2$ along [1~1~0] shown here and also along
other directions have been published previously by Noro and coworkers~\cite{Noro2002}.
In contrast to Zn and Ni substitution, Co greatly strengthens the weak ferromagnetism.
Figure~\ref{Noro}(b) shows $M_0(T)$ along [1~1~0]. The
pure Ba$_2$Cu$_3$O$_4$Cl$_2$ trace indicates that a ferromagnetic component is developed at $T \sim 330$\,K and
quickly saturates. Co substitution evidently has two effects: firstly, the transition
temperature for the appearance of the Cu$_I$ structure rises. Secondly, the
ferromagnetic moment never saturates but continues to rise steadily down to low
temperatures. A 2\% substitution triples the ferromagnetic moment with an
increase up to 4\% giving a six-fold enhancement! In Sr$_2$Cu$_3$O$_4$Cl$_2$ it was shown that the
ferromagnetic moment tracked the behavior of the order parameter for the Cu$_I$ sites~\cite{Chou1997}. Here
the order parameter behavior is shown in Fig.~\ref{Tdependence}(a). Evidently the
ferromagnetic contribution departs from this. The ferromagnetic contribution to the
magnetic structure of pure Ba$_2$Cu$_3$O$_4$Cl$_2$ is attributed to a canting of the magnetic moments. This
canting of the moments is augmented, both by Co substitution and then by
decreasing temperature.

Figure~\ref{Noro}(c) shows the magnetic susceptibility as a function of temperature and Co substitution
along the [1~1~0] direction. The overall shape changes little. The magnitude of the
susceptibility is observed to increase with Co substitution and this is the first indication
of the presence of the Co moments on the Cu$_{II}$ sites. On top of this the
susceptibility climbs strongly at the lowest temperatures with increasing Co substitution.

\section{Discussion}
\label{sec:discussion}

We have made a study of the magnetic properties of Ba$_2$Cu$_{2.95}$Co$_{0.05}$O$_4$Cl$_2$
using neutron scattering and magnetization measurements. 
For the pure material the behavior for $T < T_{N,I}$ is marked by a
single phase transition. Here we observe evidence of a new transition for $T \sim 100$~K and
a transition very similar to that of the pure material at around $T_{N,II}$. The Co substituted material also
exhibits a much larger ferromagnetic moment than the pure material.

The Cu$_I$ spins with Co substitution order in the same way as they do in the pure material, albeit with an
increased average moment. Unlike in the pure material case there is now an uncompensated moment component that is
experienced by the Cu$_{II}$ sites. There are two useful ways to think about the effect on the Cu$_{II}$ sites.
Firstly, the Co substitution can be thought of as creating new, effective, interactions between the Cu$_{II}$
spins. Secondly, the Co substitution can be thought of as longitudinal fluctuations which interact with the
thermal and quantum fluctuations of the Cu$_{II}$ spins.

\begin{figure}
\includegraphics[scale=0.8]{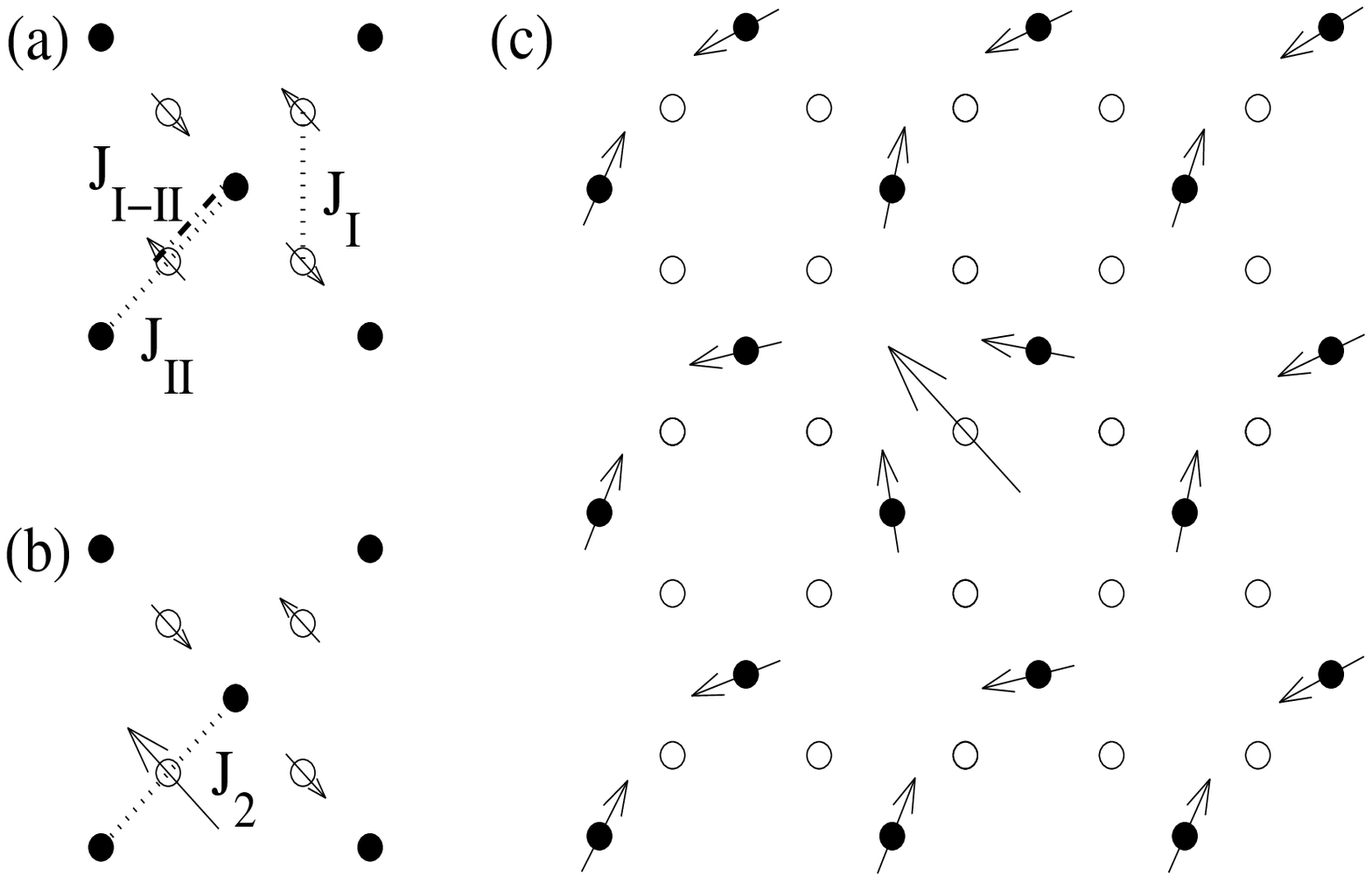} 
\caption{ \label{coupling} Schematic illustration of the magnetic couplings. Open (closed) circles
indicate Cu$_I$ (Cu$_{II}$) sites. (a) $J_I$ couples the Cu$_I$ moments and $J_{II}$ couples the
Cu$_{II}$ moments. The isotropic coupling $J_{I-II}$ between sublattices is completely
frustrated. (b) When Co is added (large arrow) there is a new effective coupling $J_2$ between
the Cu$_{II}$ moments. (c) Schematic illustration of the canted structure.
The Co moment in the center induces a ferromagnetic
ordering in nearby Cu$_{II}$ moments. Antiferromagnetic order can form independently in the
orthogonal direction.}
\end{figure}

Effective interactions between Cu$_{II}$ spins give rise to the new behavior we observe in
Ba$_2$Cu$_{2.95}$Co$_{0.05}$O$_4$Cl$_2$. Figure~\ref{coupling}(a) shows the cancellation of the Cu$_I$
contributions at the Cu$_{II}$ sites. A Co moment on the central site in this figure would strengthen the
coupling between Cu$_{II}$ moments but would not create anything new. The effective interactions arise from the
Co moments on the Cu$_I$ sites pictured in Fig.~\ref{coupling}(b). The existence of the uncompensated moment
introduces an effective ferromagnetic interaction J$_2$ between Cu$_{II}$ moments. As pointed out by Villain,
ferromagnetic interactions at random locations in an antiferromagnetic material would give rise to spin-glass
behavior in an Ising system~\cite{Villain1979}. In the case of XY or Heisenberg symmetry the frustration between ferromagnetic and
antiferromagnetic interactions can be resolved by canting of the Cu$_{II}$ moments close to the Co site. This
situation is pictured in Fig.~\ref{coupling}(c) and exhibits simultaneous ferromagnetic and antiferromagnetic
order. Villain considered a similar case with interstitial magnetic impurities: a \textit{decorated}
model~\cite{Villain1979}. 
In that case antiferromagnetic
coupling between the host and interstitial were chosen. For Ba$_2$Cu$_{2.95}$Co$_{0.05}$O$_4$Cl$_2$ the local
canted states have a non-zero 
moment as is evident in our magnetization results. In Villains model the local canting
compensates the impurity moment and a quadrupolar field results~\cite{Villain1979}.

The substituted Co moments on the Cu$_I$ sites are longitudinal fluctuations in the field experienced by the
Cu$_{II}$ moments. There is a coupling between Cu$_I$ fluctuations and the fluctuations of the Cu$_{II}$ moments.
In the pure material the coupling gives rise to collinear ordering~\cite{Shender1982, Schmalfuss2003}. 
This is because the fluctuations are
predominantly transverse to the moment direction for both Cu$_I$ and Cu$_{II}$ spins. Here the Co substituted
Cu$_I$ spins have large static longitudinal fluctuations. In order for the Cu$_{II}$ thermal and quantum
fluctuations to couple to the longitudinal fluctuations 
it is necessary for the Cu$_{II}$ spins to align perpendicular to the Cu$_I$ spins. This
model was studied by Henley~\cite{Henley1989}. 
Canting similar to Fig.~\ref{coupling}(c) is the prediction of considerations of the
effective interaction and the fluctuations. Evidently longitudinal fluctuations also exist due to the Co
substitution on the Cu$_{II}$ sublattice. These are at random fixed locations and are dilute. These fluctuations
will couple to the thermal and quantum fluctuations of the Cu$_I$ moments and hence these Co moments will also favor
perpendicular alignment.

Our elastic neutron diffraction results indicate that the Cu$_I$ and Cu$_{II}$ sublattices of
Ba$_2$Cu$_{2.95}$Co$_{0.05}$O$_4$Cl$_2$ order in a very similar manner to the pure material
(Fig.~\ref{Tdependence}). We assume that the low temperature ordering of the Cu$_{II}$ spins is collinear to the ordering of
the Cu$_I$ spins as in the pure material. This assumption is supported by the observation that substitution with Co has
hardly changed the ordering temperature and critical properties. If there were competition between collinear and
perpendicular ordering, due to the Co substitution, it might have been expected that the low temperature transition would
have moved away from Ising critical properties.
The quasi-elastic measurements (Fig.~\ref{Tdependence}(b)) 
revealed diffuse scattering at intermediate temperatures
which is associated with very short-range correlations. This scattering is not the same as that observed in the
pure material~\cite{Kim2001}. Here the correlations are due to the local
ordering close to the Co moments on both sublattices. The Co moments on the Cu$_I$ sites create a preferred
orientation for the antiferromagnetic fluctuations which is perpendicular to the Co moment. The local canted
states give rise to the short-range order. It appears likely that the drop in quasi-elastic intensity is associated with a
freezing of the perpendicular spin components while the collinear component retains the pure material behavior. 
It is the interactions between canted regions that lead to spin-glass freezing.

Other magnetic effects also occur in the same temperature range where 
we observe changes in the diffuse scattering.
The development of antiferromagnetic Cu$_I$ domains in the vicinity of 100~K has been studied in detail by Parks and
coworkers~\cite{Parks2001}. This is a rare magnetization study of antiferromagnetic domains and was possible
because of the ferromagnetic moment developed on the Cu$_{II}$ sites due to the pseudo-dipolar interaction.
Two kinds of Cu$_I$ domains are possible and the orientation of the associated ferromagnetic moments are mutually orthogonal and
so the domain populations can be studied via the magnetization signal.
The application of a high magnetic field leads to a single domain dominating the sample. As the field is reduced
domains of the other type begin to nucleate. At 100~K the susceptibility for domain wall formation is maximum.
Additional studies were made using neutron scattering showing the decrease in intensity of the (3~2~1) reflection at
around 100~K. We do not believe that the decrease in scattering intensity at the ($\frac{1}{2}$~$\frac{1}{2}$~$4$)
position at around 100~K is
associated with the proliferation of domain walls in the Cu$_I$ spin system. This is true in principle because we did
not use an applied magnetic field to create a monodomain sample. It is also true in practice 
as can be observed from our
measurements: At this temperature the Cu$_{II}$ antiferromagnetic
fluctuations at ($\frac{1}{2}$~$\frac{1}{2}$~$4$) 
are very short ranged; there is no peak in reciprocal space. By contrast, the Cu$_I$ antiferromagnetic
structure has long-range order. Evidently there is no correspondence between the size of the Cu$_I$ and Cu$_{II}$
antiferromagnetic domains at these temperatures.

Figure~\ref{Noro}(a) shows that the three regimes we have observed via neutron scattering 
are also observed in the magnetization studies presented here.
The first regime is the high temperature Cu$_I$ ordering. 
Then the spin-glass freezing transition emerges at intermediate temperatures on diluting with Co. Importantly, the
behavior in the middle regime is seen to be new: The divergence between field-cooled and zero field-cooled magnetization
measurements for the $x = 0.05, 0.1$ measurements has no parallel in the pure material (fig.~\ref{Noro}(a)).
At low temperatures the ordering of the Cu$_{II}$ sublattice persists (presumably for those moments not pinned
by a
Co moment). Figure~\ref{Noro}(b) shows that in a magnetic field applied along the
[1~1~0] direction the ferromagnetic moment M$_0$ is substantially larger than that in the pure material and has
different temperature dependence. In the pure material M$_0$ tracks the Cu$_I$ order
parameter and is induced by pseudo-dipolar coupling between sublattices. In the Co substituted system the applied
field aligns the Co moments on the Cu$_I$ sites and the canted contributions on the Cu$_{II}$ sites as pictured
in Fig.~\ref{coupling}(c). Some canting of the Cu$_I$ moments will also occur~\cite{Chou1997}.
The combination leads to a large ferromagnetic contribution. Figure~\ref{Noro}(c) shows that these local large moments also
give
an increase in the susceptibility. Magnetization measurements have been made for
Ba$_2$Cu$_3$O$_4$Cl$_2$ with the Cu diluted with Zn, Ni and Co. In the case of Zn and Ni no new
transitions appear~\cite{Noro2002} whereas with Co spin-glass behavior results. 
These differences stem from the magnitude of the
effective interactions induced. The magnetic interaction between the spin of 
a Cu$^{2+}$ ion on a Cu$_I$ site and the Cu$_{II}$ spins 
is canceled by the other spins
on the Cu$_{I}$ sublattice. Ni$^{2+}$ and Zn$^{2+}$ have moments that differ from Cu$^{2+}$ by $\frac{1}{2}$.
Hence Zn and Ni on Cu$_I$ sites result in an uncompensated spin one-half moment which is
apparent to the Cu$_{II}$ moments. Co$^{2+}$ has a moment that differs from Cu$^{2+}$ by 1 hence 
for Co the effect is doubled. Since J$_{II}$ and J$_{I-II}$ are equal and
opposite only Co is likely to give rise to new and dominant interactions between the Cu$_{II}$ spins.

Our neutron scattering and magnetization results are consistent with the Co substitution leading to a new transition in the
Cu$_{II}$ spin system. Spin-glass freezing at $\sim$ 100~K precedes the antiferromagnetic ordering at $\sim$ 30~K. We
interpret our observations in-terms of the involvement of both the parallel and perpendicular spin components on the
Cu$_{II}$ sites (Fig.~\ref{coupling}(c)). The change in behavior is due to the influence of the Co moments located 
on the Cu$_I$
sites; this can be explained in terms of new effective interactions or alternatively coupling between fluctuations.
Either way the
Cu$_{II}$ moments are caused to form local canted states around the Co moments. The local canted states
frustrate each other and lead to spin-glass freezing. The local cluster has a substantial
moment and this is reflected in the magnetization measurements. At lower temperatures the portion of the
Cu$_{II}$ moment that is collinear with the Cu$_I$ ordering direction (and not pinned by a Co moment) 
orders antiferromagnetically as in the pure
material. This system exhibits a delicate balance of quenched random and quantum fluctuations. The former
controls the magnetic behavior at intermediate temperatures while the latter yields order at lower temperatures.

\section{Acknowledgments}

We are grateful to P.M.~Gehring for invaluable help with the neutron scattering measurements and to 
Y.J. Kim for sharing expertise concerning the properties of pure \textit{2342}
materials. We acknowledge the support of the National Institute of Standards and Technology, US.
Department of Commerce, in providing the neutron research facilities used in this
work. Funding in Toronto was provided by the Natural Science and Engineering Research
Council and in Edinburgh by the EPSRC (Grant GR/S10377/01).

\end{document}